\documentclass[aps,pra,twocolumn,superscriptaddress,showpacs]{revtex4}
\pdfoutput=1 

\usepackage[english]{babel}
\usepackage{graphicx}
\usepackage{amsmath}
\usepackage{amssymb}
\usepackage{color}
\usepackage{cancel}

\newcommand{\tm}{(\text{th})}
\newcommand{\Di}{\mathcal{D}}
\newcommand{\Ki}{\mathcal{K}}
\newcommand{\Bi}{\mathcal{B}}

\newcommand{\Cor}{\mathfrak{C}}
\newcommand{\Dsla}{\cancel{\mathcal{D}}}

\newcommand{\trace}{\text{Tr}}

\begin{document}

\title{Entanglement renormalization and boundary critical phenomena}

\author{P. Silvi}
\affiliation{International School for Advanced Studies (SISSA),
  		Via Beirut 2-4, I-34014 Trieste, Italy}
\author{V. Giovannetti}
\affiliation {NEST,  Scuola Normale Superiore \& CNR-INFM, Piazza
		dei Cavalieri 7, I-56126 Pisa, Italy}
\author{P. Calabrese}
\affiliation{Dipartimento di Fisica dell' Universit\'{a} di Pisa and INFN, 56127 Pisa, Italy }
\author{G. E. Santoro}
\affiliation{International School for Advanced Studies (SISSA),
  		Via Beirut 2-4, I-34014 Trieste, Italy}
\affiliation{CNR-INFM Democritos National Simulation Center, Via Beirut 2-4, I-34014 Trieste, Italy}
\affiliation{International Centre for Theoretical Physics (ICTP), P.O.Box 586, I-34014 Trieste, Italy }
\author{R. Fazio}
\affiliation {NEST,  Scuola Normale Superiore \& CNR-INFM, Piazza
		dei Cavalieri 7, I-56126 Pisa, Italy}
\affiliation{Center for Quantum Technologies, National University of Singapore, Republic of Singapore}

\date{\today}

\begin{abstract}
The multiscale entanglement renormalization ansatz is applied to the study of boundary critical phenomena. We compute averages of local operators as a function of the distance from the boundary and the surface contribution to the ground state energy.  Furthermore, assuming a uniform tensor structure, we show that the multiscale entanglement renormalization ansatz implies an exact relation between bulk and boundary  critical exponents known to exist for boundary critical systems.

\end{abstract}

\pacs{03.67.-a,05.30.-d,89.70.-a}

\maketitle

\section{Introduction} 
\label{sec:intro}
Variational approaches based on tensor networks~\cite{CV} are a novel powerful numerical tool believed to be the key
ingredient to simulate efficiently quantum-many body systems. Although a detailed understanding  of their potentialities and their 
limitations is presently under scrutiny, there are already a number of encouraging results.  Variational Tensor Networks (VTN) are free 
of most of the problems of traditional numerical methods. Differently from quantum Monte Carlo methods, VTN do not suffer of the sign 
problem. Compared to density matrix renormalization group~\cite{DMRG},  they are more versatile and allow to simulate efficiently critical 
correlations, long-range interactions and two- and higher-dimensional quantum systems. Indeed the density matrix renormalization group can be  
reformulated in terms of  a particular class of tensor networks known as Matrix Product States~\cite{mps}. VTN include also projected 
entangled pair states~\cite{peps} that generalize matrix product states  in dimensions higher than one and  weighted graph states~\cite{wgs} 
designed to study systems with long-range interactions. 

Among the proposed VTNs, a very promising one is the so-called Multiscale Entanglement Renormalization Ansatz (MERA)~\cite{mera}.  
MERA has been already applied successfully  to the study of  a number of  different physically relevant systems, like quantum models on a 
two-dimensional lattices~\cite{cincio, Evenbly:2009p1759},  interacting fermions~\cite{corboz}, and critical 
systems~\cite{jova08,pev-09,Giovannetti:2009p1563,Montangero:2008p1565}, only to mention a few of the most remarkable examples. 
The capability of  MERA to describe accurately critical systems derives directly from the scale-invariant self similarity of its tensor structure, 
intimately related to  a real-space  renormalization procedure. The structure of the MERA state is designed~\cite{mera} in such a way 
to reproduce scale-invariance and so, in one-dimensional systems, it naturally encoded several important features of the Conformal Field
 Theory (CFT) underlying  the critical lattice model~\cite{pev-09}. 
 The critical exponents can be computed directly from the spectrum of the MERA transfer matrix~\cite{jova08}. 

Critical systems can however lack translational invariance, due to the presence of a physical boundary or to an impurity.  The study of 
boundary critical phenomena is, since many years, a very active field of research which ranges from the study of the critical magnets with 
surfaces to quantum impurity problems  (as e.g. Kondo) or the Casimir effects (for a review of the field see for example~\cite{bcs}).
The presence of the boundary does not spoil conformal invariance. Oppositely, boundary CFTs have a very rich structure and a deep 
mathematical foundation (see e.g.~\cite{cardy-05}).   
While in fact it started only as the study of critical two-dimensional systems in system with boundaries (surface critical behavior), it 
found applications to open-string theory (D-branes), quantum impurity problems \cite{s-98}, 
quantum out-of-equilibrium studies (quantum quenches \cite{cc-q}) just to cite a few.
Furthermore, it attracted a large attention from the mathematical community for the recent 
developments of stochastic Loewner evolution \cite{sle}.

In view of the connection between MERA and CFT, it is natural to wonder whether  the MERA tensor-network can be employed to study 
quantum systems with boundaries. In this work we introduce and analyze an entanglement renormalization tensor-network design which 
takes into account the presence  of a (critical) boundary, and we study its properties.
This is implemented by  allowing the edge spin at the boundary of the system  to interact at each step with an \emph{ancillary} 
 element, describing a fictitious degree of freedom.  
Similarly to Refs.~\cite{jova08,Giovannetti:2009p1563,pev-09},  we will focus on 
homogeneous configurations,  where tensor elements of the same class are also identical to each other. 
Interestingly enough  this ansatz is able to capture  some important properties  predicted by boundary CFT. 
Specifically, we show that  the critical exponent associated to the decay of any 
one-point function (as function of the distance from the  boundary) is always half of the one of the bulk two-point correlation function 
corresponding to the same scaling operator.  We also compute the boundary corrections to the ground state energy.

%%%%%%%%%%%%%%%%%%%%%%%%%%%%%
\begin{figure*}[htb]
\hspace*{-0.6cm}
 \includegraphics[width=280pt]{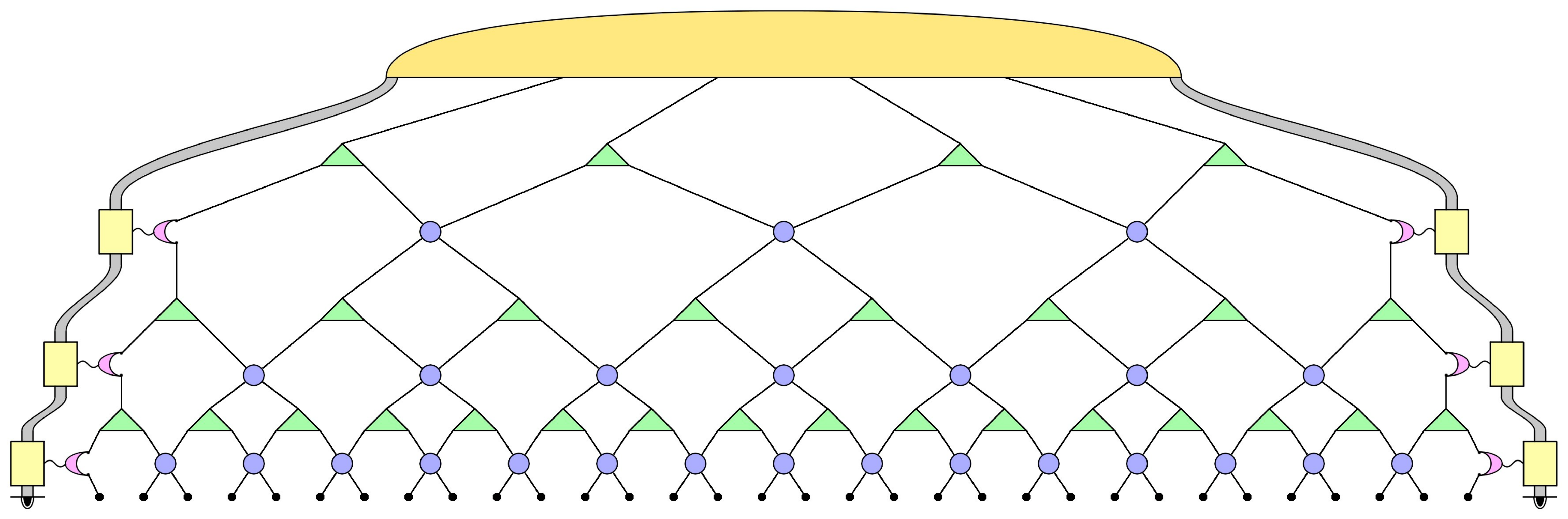} \hspace*{1.cm}
 \includegraphics[width=140pt]{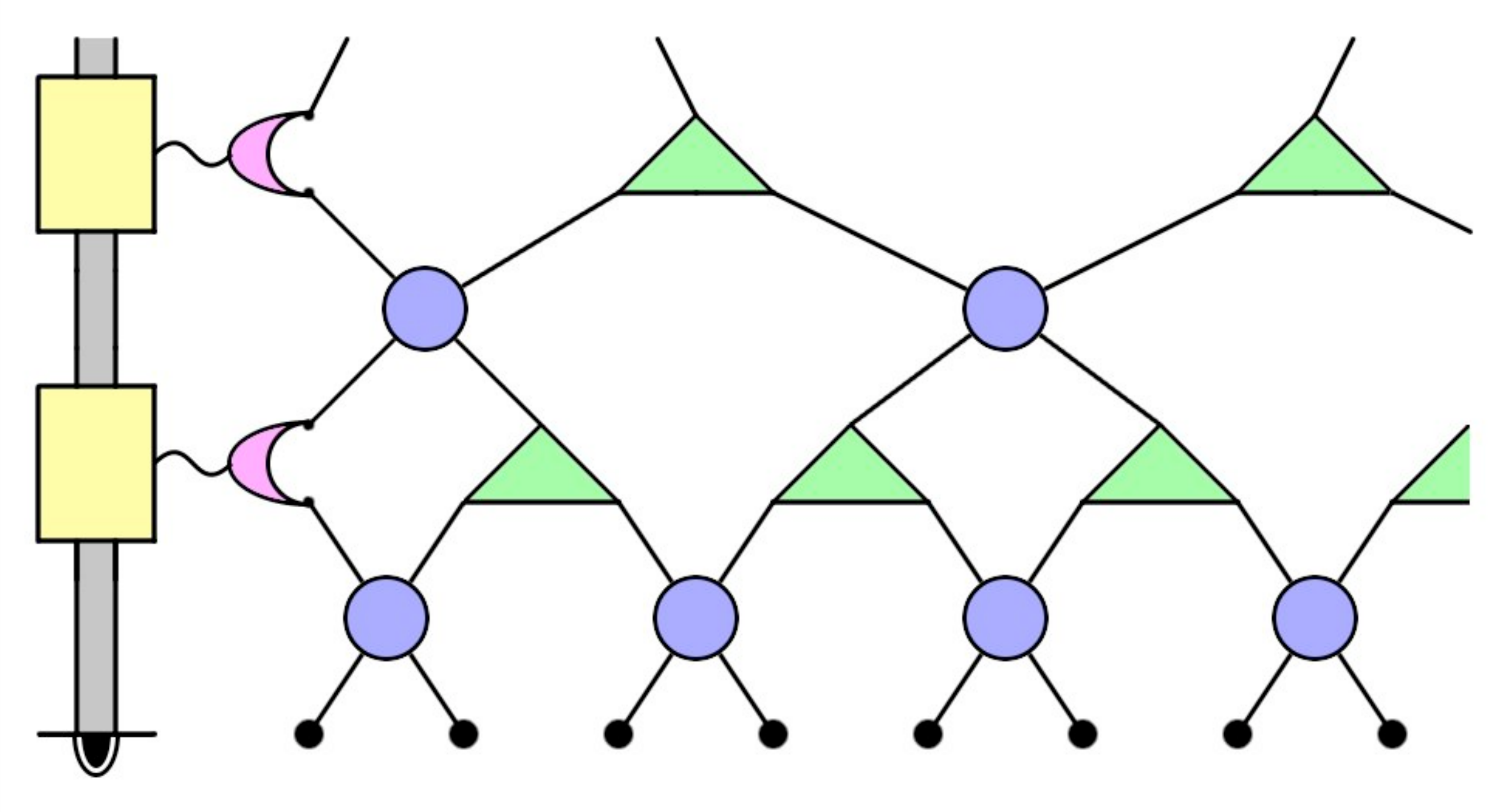}
\caption{ \label{fig:chibera}
Left panel - Entanglement renormalization network representation for $N=32$ sites. In the bulk the MERA structure consists of 
isometries and disentanglers. The boundary is represented by an additional ancillary system indicated by the grey stripe.
Right panel - Alternative insertion of the boundary obtained  by applying the ancilla  interaction tensors (magenta crescent-moons) at the
same level of the disentanglers instead of the renormalizers, as done in the left panel. }
\end{figure*}
%%%%%%%%%%%%%%%%%%%%%%%%%%%%%%%%%%%%%%%%

The paper is organized as follows. In Sec.~\ref{sec:ans} we introduce the tensor network and its main properties. In Sec.~\ref{sec:ex} 
we discuss how the expectation values of local observable can be computed. Assuming a uniform tensor structure, 
in the thermodynamic limit,  these expectation values decay as power law. We relate the associated critical exponents to those of the 
corresponding two-point correlations in the bulk.  In Sec.~\ref{sec:tra} we discuss the boundary corrections to the ground state energy. 
The conclusions of our work are summarized in Sec.~\ref{sec:con}.

\section{The tensor network} 
\label{sec:ans}

Consider a 1D lattice of $N = 2^{n+2}$ spins (sites), of a given local dimension $d$, with open boundary conditions. 
A generic pure state  of such system can  be expressed as 
\begin{eqnarray}
|\psi^{(n)}\rangle = \sum_{\ell_1, \ldots , \ell_N = 1}^{d} \mathcal{T}_{\ell_1, \ldots \ell_N}
 |\xi_{\ell_1} \ldots \xi_{\ell_N} \rangle \label{eq1}\;,
 \end{eqnarray} 
with $\{ |\xi_{i}\rangle \}_i$ a canonical basis for the single qudit and $\mathcal{T}$ a  
type-$ \mbox{\tiny{$\left(\begin{array}{c} 0 \\N \end{array}\right)$}}$ 
tensor.  Following the prescriptions of the MERA structure~\cite{mera}, we assume a formal decomposition of $\mathcal{T}$ which is 
schematically sketched in Fig.~\ref{fig:chibera}. Here we use the standard graphical convention (see e.g. Ref.~\cite{mera}) for which
each node of the graph represents a tensor (the emerging legs of the node being its indices), while a link connecting any two nodes represents 
the contraction of the corresponding indices. 
The yellow element on the top of the figure describes a   
type-$\mbox{\tiny{$\left(\begin{array}{c} 0 \\6 \end{array}\right)$}}$ tensor ${\cal C}$   of elements ${\cal C}_{a, \ell_1,\ell_2,\cdots,  \ell_4,a'}$,  
that we can call {\em hat} tensor. 
The green triangles represent instead the same $d \times d^2$  renormalizer tensor $\lambda$ of 
type-$ \mbox{\tiny{$\left(\begin{array}{c} 1 \\2 \end{array}\right)$}}$ of elements 
${\lambda}^{u}_{\ell_1,\ell_2}$, and the blue circular elements represent  the same $d^2\times d^2$ disentangler tensor $\chi$ of  type-$
\mbox{\tiny{$\left(\begin{array}{c} 2 \\2  \end{array}\right)$}}$ of elements $\chi^{u_1,u_2}_{\ell_1,\ell_2}$.

At the boundary, we introduce extra tensors that couple the  sites at the border  with an ancillary degree of freedom represented in Fig. 
\ref{fig:chibera} by the thick grey strip. 
These new elements  form the lateral edges of the  network  and describe the boundary at each level of the MERA, i.e. at each level of 
the renormalization flow.  As shown in the figure, each of them can  be viewed as a matrix product state  (yellow squares) whose bonding dimensions  coincide with the  coordinate space of the ancillas, which is coupled to the bulk via  local coupling-elements (drawn as magenta 
crescent-moons in the figure).
Via purification, we can always choose the dimension of such ancilla  to be large enough so that the resulting interaction is described by a 
unitary operator, 
that we indicate as $\alpha$,  a type-$
\mbox{\tiny{$\left(\begin{array}{c} 2 \\2  \end{array}\right)$}}$ tensor of elements $\alpha^{u_1,u_2}_{\ell_1,\ell_2}$.
Similarly to the case of the $\lambda$s and of the $\chi$s, we will also assume these  elements to be uniform in the network (possibly allowing the 
ones on the left-hand-side of the structure to differ from the ones on the right-hand-side  \cite{footbcco}). 

\begin{figure}
 \centering
 \includegraphics[width=230pt]{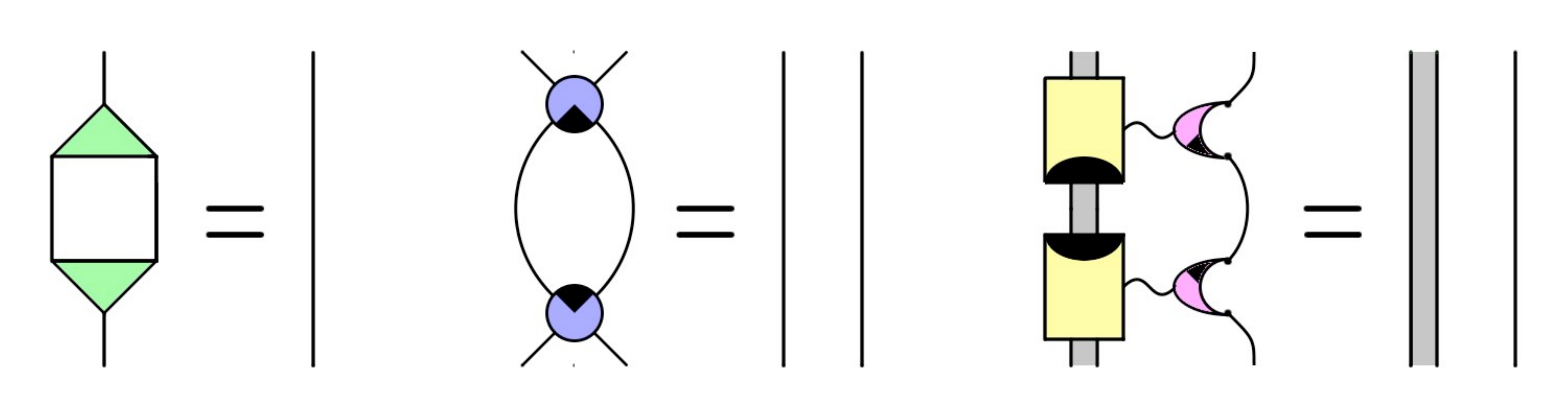}
 \caption{ \label{fig:rulez}
 Rules for the elements of the tensor network of Fig.~\ref{fig:chibera}. The black marks drawn upon the tensors
 are used to make distinction between top and bottom. The first two schemes correspond to the the contractions of Eq.~(\ref{ffd}),
while the last one  represents the unitary constraint for the edge elements of the tensor.
 }
\end{figure}

As customary with MERA-like configurations, to enforce efficient evaluation of local observables and correlation functions, the various elements 
of the network are assumed to obey specific contraction rules (a detailed analysis of the efficiency requirements for MERA  can be
found in Refs.~\cite{mera,EVE,Giovannetti:2009p1563}). In particular the renormalizers and the disentanglers obey isometric and unitary 
constraints respectively, i.e. 
\begin{eqnarray} \label{ffd}
\sum_{k_1,k_2} \lambda^{u}_{k_1,k_2} \bar{\lambda}_{\ell}^{k_1,k_2} = \delta^{u}_{\ell}, \quad
\sum_{k_1,k_2} \chi^{u_1,u_2}_{k_1,k_2} \bar{\chi}_{\ell_1,\ell_2}^{k_1,k_2} = \delta^{u_1}_{\ell_1}\; \delta^{u_2}_{\ell_2},
\end{eqnarray}
where $\delta^{u}_{\ell}$ is the Kronecker delta, while $\bar{\lambda}_{\ell}^{u_1,u_2} \equiv  ({\lambda}^{\ell}_{u_1,u_2})^*$ 
and  $\bar{\chi}_{\ell_1,\ell_2}^{u_1,u_2} \equiv  ({\chi}^{\ell_1,\ell_2}_{u_1,u_2})^*$  are the adjoint counterparts of the  $\lambda$ and $\chi$ 
tensors respectively, obtained by exchanging their lower and upper indices and taking the complex conjugate. 
Similar conditions are imposed also for the edge tensors 
\begin{eqnarray} \label{ffe}
\sum_{k_1,k_2} \alpha^{u_1,u_2}_{k_1,k_2} \bar{\alpha}_{\ell_1,\ell_2}^{k_1,k_2} = \delta^{u_1}_{\ell_1}\; \delta^{u_2}_{\ell_2}.
\end{eqnarray}
These rules are graphically represented in  Fig.~\ref{fig:rulez}.
Finally to ensure proper state normalization, the tensor ${\cal C}$ is supposed to satisfy the 
identity $\sum_{a, k_1, \cdots, k_4, a'}{\cal C}_{a, k_1, \cdots,k_4, a'} \bar{\cal C}^{a, k_1, \cdots, k_4, a'} =1$.

It is worth noticing that, by  simply re-arranging the various tensorial components, 
an entanglement renormalization configuration  which differs from the one given in Fig.~\ref{fig:chibera} can be obtained. 
In fact,  the ancilla interaction tensors $\alpha$ (the magenta crescents) can be applied
at the same level of the disentanglers, instead of the renormalizers  (see Fig.~\ref{fig:chibera} right panel).
Despite their different appearance, it can be shown that these two structures are formally equivalent to each other. 
This can be verified by grouping together the edge-ancilla interaction with the nearest linked element belonging to the lower half-level.
Now, by just performing a polar decomposition~\cite{HORN}, we obtain a structure of the opposite type (although, the very first spin
of the chain is now taken out the system and put into the ancilla, while the second one becomes the edge spin).
Having acknowledged this equivalence, in the rest of the paper we will work with  the  structure of the left panel of Fig.~\ref{fig:chibera}.

%%%%%%%%%%%%%%%%%%%%%%%%%%%%
%%%%%%%%%%%%%%%%%%%%%%%%%%%%%%%%%%%%%%%%%%

\section{Local averages in the presence of a boundary}
 \label{sec:ex}

In the presence of the boundary, the average of any local observable depends on the distance from the boundary itself. 
For a one-dimensional critical system, the case we consider here, the space-dependence will be a power-law characterized 
by a set of critical exponents. In this Section we show how to compute local averages and how to extract these 
exponents. 

Consider  a  family ${\cal F}\equiv \{ |\psi^{(n)}\rangle ; n =2, 3, \cdots \}$ of  states $|\psi^{(n)}\rangle$ of increasing sizes, 
described by  homogeneous networks of the form shown in  Fig.~\ref{fig:chibera}, each sharing the same structural elements 
(renormalizer, disentangler, edge-ancilla interaction, hat).  For such family  we  want to calculate the expectation value 
of a general observable acting on a small group of neighboring sites located at a given distance from the closest edge of the system,
say the left one. For instance, in the case of a three-sited observable $\Theta_\ell$ acting on the sites  $\ell,\ell+1$, and $\ell+2$~\cite{MOTA}
we have
\begin{equation} \label{eq:prima}
 \langle \Theta_\ell \rangle^{(n)} \equiv 
 \langle \psi^{(n)} | \Theta_\ell | \psi^{(n)} \rangle =
 \trace [ \Theta_\ell  \cdot \rho^{(n)}_{\ell,\ell+1,\ell+2}],
\end{equation}
where the site indices are counted starting from the leftmost spin
as the first one, and where $\rho^{(n)}_{\ell,\ell+1,\ell+2}$ is the reduced density matrix of $|\psi^{(n)}\rangle$ associated with the selected spins.

\begin{figure}[!b]
 \centering
 \includegraphics[width=230pt]{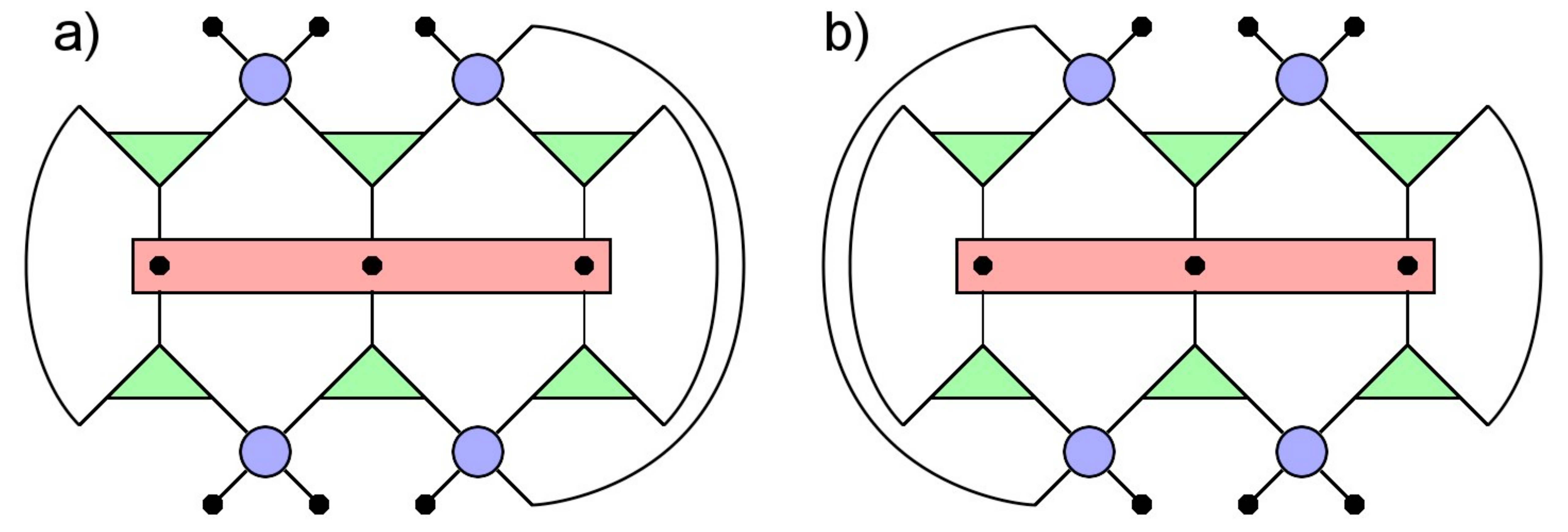}
 \caption{ \label{fig:dielledierre}
 Sketch of the CPT maps \emph{a)} $\Di_L$ and \emph{b)} $\Di_R$; the contracted tensors picture represents
 their Kraus decomposition.}
\end{figure}

\begin{figure}
 \centering
 \includegraphics[width=110pt]{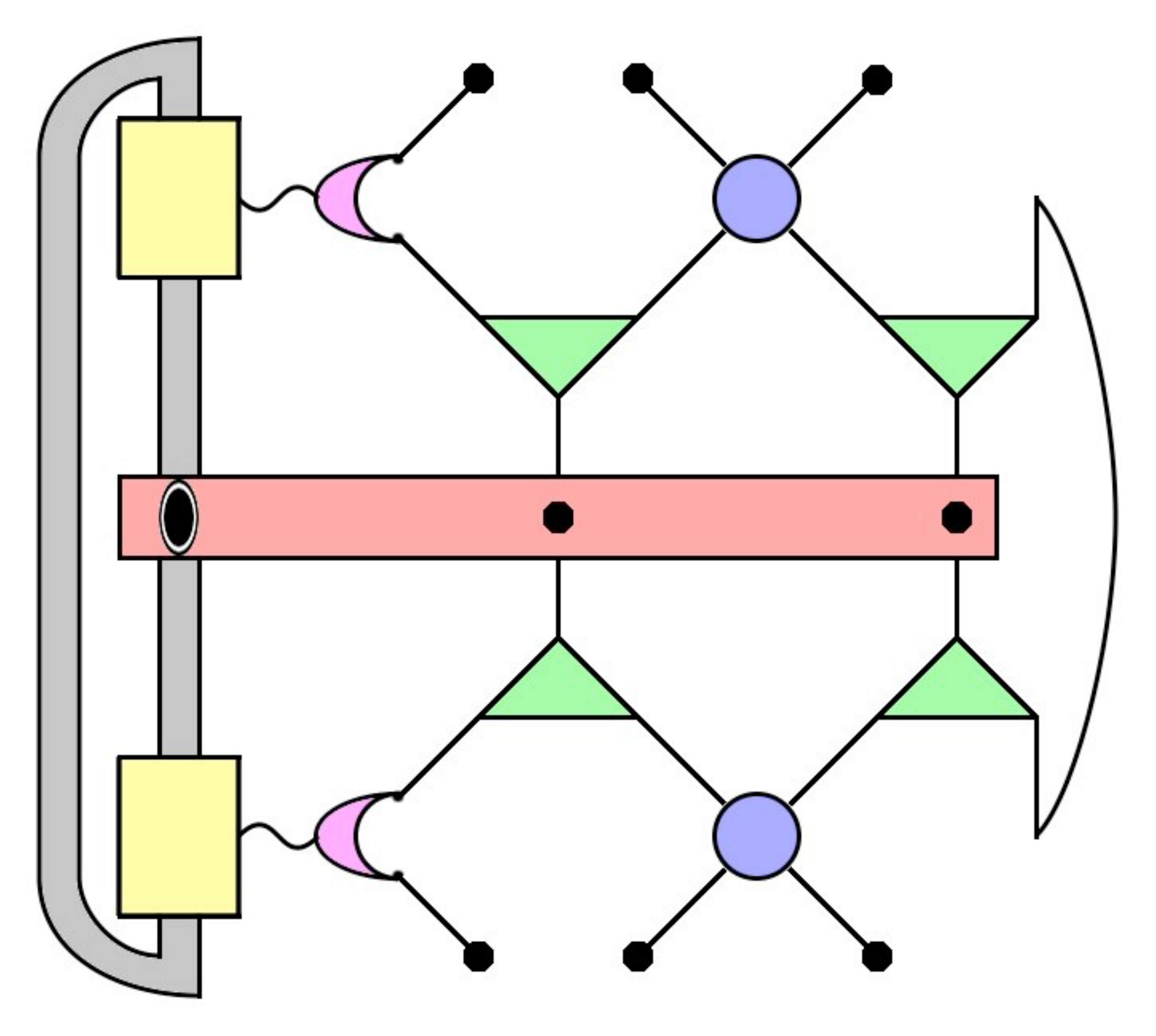} \hspace{2mm}
  \includegraphics[width=110pt]{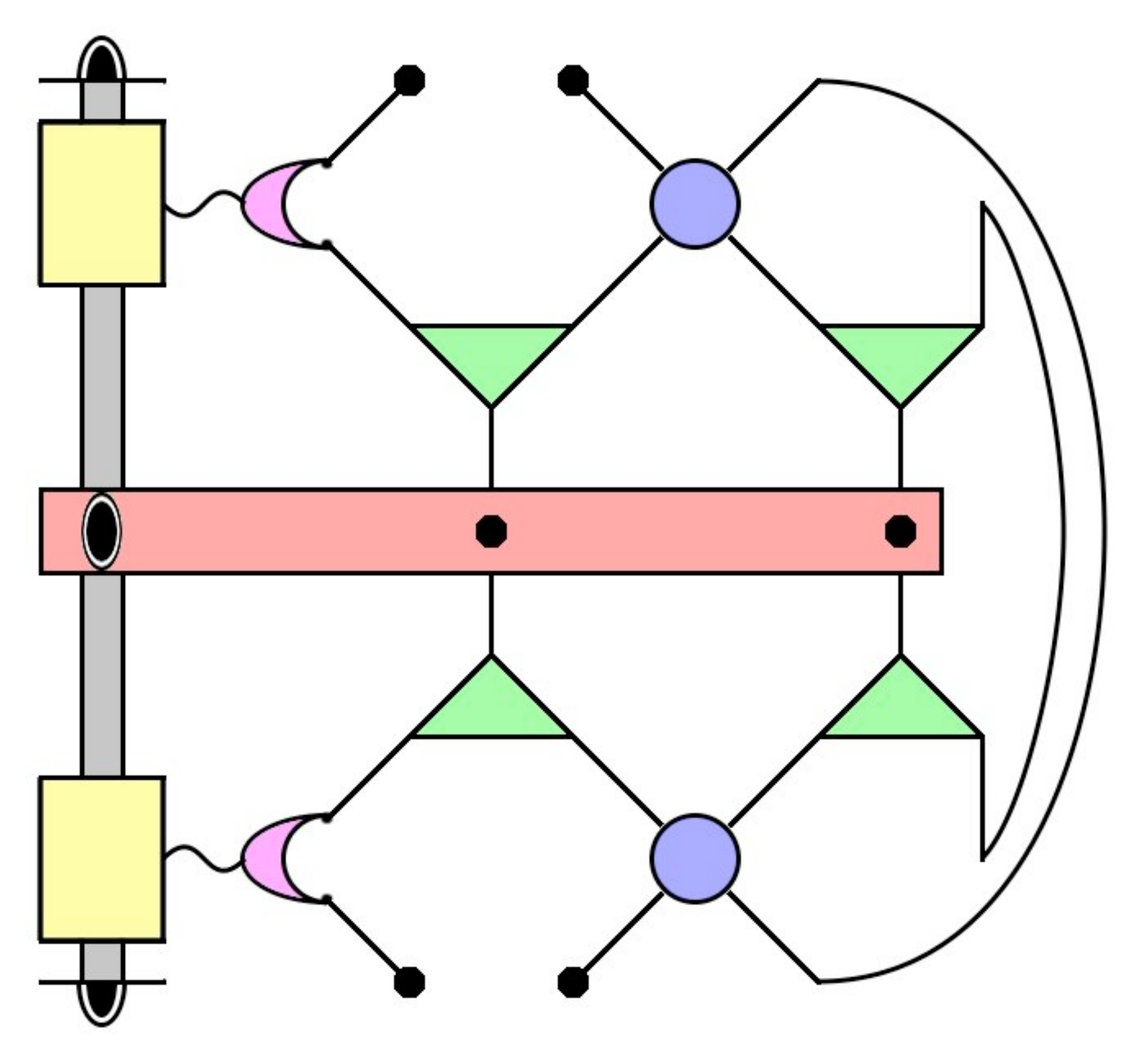}
 \caption{ \label{fig:kel}
Tensor graph representations of the CPT map $\Ki_L$ (left) and $\Bi_L$ (right) defined in  Eq.~(\ref{mappak}) and (\ref{bielledef}) respectively.}
\end{figure}

We assume a uniform MERA structure. This assumption may seem an over-simplification for a system which is not translational invariant, but 
it turns out that it naturally accounts for the underlying (boundary) conformal invariance. 
From the locality requirements imposed in Fig.~\ref{fig:rulez}, it is straightforward to verify that for all  $1\leqslant \ell \leqslant 2^{n+1}-2$ 
and  $n\geqslant 1$,
the following  recursion rules apply: 
\begin{equation} \label{eq:ricors}
 \begin{array}{c}
 \rho^{(n)}_{2\ell,2\ell+1,2\ell+2} = \Di_L \left( \rho^{(n-1)}_{\ell,\ell+1,\ell+2} \right)\;, \\ \\
 \rho^{(n)}_{2\ell+1,2\ell+2,2\ell+3} = \Di_R \left( \rho^{(n-1)}_{\ell,\ell+1,\ell+2} \right)\;,
 \end{array}
\end{equation}
where $\rho^{(n-1)}_{\ell,\ell+1,\ell+2}$ and $\rho^{(n-1)}_{\ell,\ell+1,\ell+2}$ are 3-sites reduced density matrices  of  
$|\psi^{(n-1)}\rangle \in {\cal F}$. In these expressions, $\Di_L$ and $\Di_R$ are completely positive trace preserving (CPT) maps 
that  depend only on the bulk elements of the network 
(indeed they coincide with the $\Di_L$ and $\Di_R$ maps of an ordinary
homogenous MERA  with the same $\lambda$ and $\chi$~\cite{jova08}) and whose
 formal expression is graphically depicted in Fig.~\ref{fig:dielledierre}.
 %%%%%%%%%%%%%%%%%%%%%%%%%%%%
%%%%%%%%%%%%%%%%%%%%%%%%%%%%%%%%%%%%%%%%%%
By means of  the  the renormalization procedure implied by Eq.~\eqref{eq:ricors},  at each application of the map the 
site over which the average is performed  approaches  the boundary in an exponential fashion. 
Correspondingly the network depth decreases linearly. Upon reaching the boundary one has to define further operations:
\begin{equation} \label{mappak}
 \rho^{(n)}_{1,2,3} = \mathcal{K}_L \left( \rho^{(n-1)}_{A,1,2} \right)\;,
\end{equation}
where $A$ refers to the degree of freedom belonging to the ancillary
system, and $\Ki_L$ is again a CPT map, sketched in Fig.~\ref{fig:kel} (left).
At this point the causal cone of the ancilla jointed with the first two sites
is stable. Indeed one has
\begin{equation}
 \rho^{(n)}_{A,1,2} = \Bi_L \left( \rho^{(n-1)}_{A,1,2} \right),
 \label{bielledef}
\end{equation}
where $\Bi_L$ is the CPT map represented in Fig.~\ref{fig:kel} (right).
Analogously to $\Bi_L$ and $ \mathcal{K}_L$, we define the CPT maps at the right boundary from the mirror images of 
Fig. \ref{fig:kel} and we call them $\Bi_R$ and $ \mathcal{K}_R$ respectively.

 Because of the stability of the causal cone~\cite{mera,EVE,Giovannetti:2009p1563}, approaching the thermodynamical limit
we can determine the reduced density matrix in proximity of the boundary by
calculating the fixed point of $\Bi_L$. 
This is  unique provided that the CPT map is mixing~\cite{NJP,jova08}, i.e. 
\begin{equation}
 \lim_{n \to \infty}\rho^{(n)}_{A,1,2} = \rho^{f}_{A,1,2} = \Bi_L \left( \rho^{f}_{A,1,2} \right).
\end{equation}
We can now use this argument to obtain the expectation value in Eq. \eqref{eq:prima} for infinite volume. 
The resulting expression becomes particularly simple when
$\Di_L = \Di_R = \Di$.  In this case indicating  the integer part of $\log_2 \ell$ with $\lfloor \log_2 \ell\rfloor$,
we have 
\begin{equation} \label{eq:sviluppo}
 \langle \Theta_{\ell} \rangle^{(\infty)} = 
 \trace \left[ \Theta \cdot \Di^{\lfloor \log_2 \ell\rfloor }\circ \Ki_L (\rho^{f}_{A,1,2}) \right]\;, 
\end{equation}
where  ``$\circ$" stands for super-operator composition and where 
${\cal D}^\tau$ 
describe a $\tau$ reiterated applications of the map ${\cal D}$.

We can now exploit the Jordan block decomposition~\cite{HORN}  to simplify further this expression.
%%%%%%%%%%%%%%%%%%%%%%%%%%%%%%%%%%%%%%%%%%%%%
%%%%%%%%%%%%%%%%%%%%%%%%%%%%%%%%%%%%%%%%%%%%%%
Adapting the derivation for the bulk in Ref. ~\cite{Giovannetti:2009p1563} to the boundary case,  
we easily get 
\begin{equation} \label{ddd}
  \langle \Theta_{\ell} \rangle^{(\infty)} = 
 \sum_{\kappa}  \kappa^{\lfloor \log_2 \ell\rfloor}
\; {g}_\kappa(\lfloor \log_2 \ell\rfloor) 
 \;,
 \end{equation}
where the sum spans over the eigenvalues $\kappa$ of $\Di$ and  ${g}_\kappa(\cdot)$ 
 is a (finite degree) polynomial in its main argument with coefficients
which depends on $\rho^{f}_{A,1,2}$ and $\Di$.
Since CPT maps are contractive, the  values of $\kappa$ entering in Eq.~(\ref{ddd}) belong to the unit circle (i.e. $|\kappa|\leqslant 1$).
Furthermore, if  $\Di$ is mixing (which is a reasonable assumption~\cite{jova08}) then its spectrum admits a unique eigenvector  (the fix point $\rho^f$ of the channel) associated with $\kappa=1$; all the remaining eigenvalues have modulus strictly smaller than $1$. Under these circumstances, 
in the limit of large  distance from the boundary, the quantity $\langle \Theta_{\ell} \rangle^{(\infty)}$ converges toward its bulk limit which 
is obtained by computing the expectation value of $\Theta$ on the fix point of the channel, i.e.  $\langle \Theta_{\text{bulk}} \rangle^{(\infty)} = \mbox{Tr}[ \Theta \rho_f]$. 
The deviations from such limiting expression can be  evaluated by keeping the largest contribution associated with the terms with $\kappa \neq 1$. 
This yields an  
 exponential decay in $\log_2 \ell$ of the form
\begin{eqnarray} \label{befo}
 \langle \Theta_{\ell 
 } \rangle^{(\infty)} &=& \langle \Theta_{\text{bulk}} \rangle^{(\infty)} + 
 \sum_{\kappa\neq 1}  \kappa^{\lfloor \log_2 \ell\rfloor}
\;  {g}_\kappa(\lfloor \log_2 \ell\rfloor) \nonumber \\
 &=& \langle \Theta_{\text{bulk}} \rangle^{(\infty)}+    \ell^{\log_2 |\bar{\kappa} |}   \; g'(\log_2 \ell) \;,
\end{eqnarray}
where $\bar{\kappa}$ is the  eigenvalue of $\Di$ which has the largest absolute value smaller than one and  
which contributes not trivially to Eq.~(\ref{ddd}),  and where $g'(\log_2 \ell)$ is instead
some complicated function which is dominated by a polynomial of $\log_2 \ell$. 
In particular, if $\Theta$ is an eigenvector   of the Heisenberg adjoint  of the channel $\Di$, then 
Eq.~(\ref{befo}) yields an exact power-law decay (without polynomial corrections), i.e. 
\begin{eqnarray} \label{befo1}
 \langle \Theta_{\ell 
 } \rangle^{(\infty)} &=&c \; \kappa^{\lfloor \log_2 \ell\rfloor}  \;, 
\end{eqnarray}
where $c =  \;  \mbox{Tr} [ \Theta \; {\cal K}_L  (\rho_{A,1,2}^f) ]$, and  where $\kappa$ is the associated  eigenvalue (notice that 
for such a $\Theta$ one has $\langle \Theta_{\text{bulk}} \rangle^{(\infty)}=0$). 
The above expressions 
show  that  the quantities $- \log_2 |{\kappa}|$ play the role of the critical exponents of the system.   
It is now evident that these critical exponents are  the half of the corresponding ones for two-point correlation functions computed in the bulk, 
a well-known result in  conformal field theory \cite{cardy-05}. 
For instance  fixing the distance $\Delta \ell = 2^m$  of the two points,  the bulk connected correlation function has  been computed~\cite{jova08}, and it holds 
\begin{eqnarray}
\Cor^{(n)}_{\Delta \ell} \equiv \frac{1}{N} \sum_{\ell = 1}^{N}
  [ \langle \Theta_{\ell} \, \Theta_{\ell + \Delta\ell} \rangle^{(n)} -
  \langle \Theta_{\ell} \rangle^{(n)} \langle \Theta_{\ell + \Delta\ell}
  \vphantom{\sum} \rangle^{(n)} ] \nonumber \\= \mbox{Tr} [ (\Theta \otimes \Theta) \; \Dsla^{\log_2 \Delta \ell } (\sigma) ] =  \sum_{\xi\neq 1} \xi^{\log_2 \Delta \ell} \; h_\xi(\log_2 \Delta\ell), 
  \end{eqnarray}
where the summation is performed over the eigenvalues $\xi$ of the CPT map 
 $\Dsla = \frac{1}{2} \left(
 \Di_L \otimes \Di_L + \Di_R \otimes \Di_R \right)$, and where $h_\xi(\cdot)$ is  a polynomial function of its argument (in this expression
 $\sigma$ stands for a traceless operator while finally $N=2^{n+2}$ is the size of the associated homogeneous  MERA with periodic boundary conditions).
 The result then follows by noticing that by construction  $\Dsla = \Di \otimes\Di$ so that the $\xi$ can be expressed as products $\kappa \kappa'$ of the eigenvalues of $\Di$. 
 In particular if as in Eq.~(\ref{befo1}),
 $\Theta$ is an eigenvector the Heisenberg adjoint of $\Di$ we have that $\Theta \otimes \Theta$ is an eigenvector of the adjoint of $\Dsla$ at the eigenvalue  $\xi = \kappa^2$ and thus,
 \begin{eqnarray}
\Cor^{(n)}_{\Delta \ell}  =  c' \kappa^{2 \log_2 \Delta \ell}  = 
c' 
\;(\Delta \ell)^{2 \log_2 \kappa } \;,
  \end{eqnarray}
which proves the claim (here  $c' = \mbox{Tr} [(\Theta \otimes \Theta) \sigma]$).

\section{Boundary contribution to the ground state energy} 
\label{sec:tra}

In the presence of a boundary, the average of extensive observables (the ground state energy for example) does 
contain a bulk and a boundary contribution (negligible in the thermodynamic limit).
In this section, we evaluate the {\it density} of the ground-state energy
for a local Hamiltonian (with interactions among sites at maximum distance $\nu$) of the form
\begin{equation}
 \mathcal{H} = \frac{1}{L-\nu+1} \sum_{j = 1}^{L - \nu+1} H_{j \ldots j+\nu},
\end{equation}
where $\nu$ is the number of sites involved in the model interaction $H$.
While this problem is easily solved in a level-recursive manner for
a MERA structure (in which periodic boundary conditions hold), when
explicit conditions over a defined boundary are involved things become slightly
more complicated. 

Suppose, for simplicity, that the interaction is again a $\nu =$ 3-body operator,
therefore
\begin{equation}
 \langle \mathcal{H} \rangle = \trace [ H_3 \cdot \bar{\rho}_3^{(n)}]\;,
\end{equation}
where
\[
 \bar{\rho}_3^{(n)} \equiv \frac{1}{2^{n+2}-2} \sum_{j = 1}^{2^n} \rho^{(n)}_{j,j+1,j+2}\;\; .
\]
We need to build a recursive function which relates this average density matrix
to the one belonging to the previous tensor level $\bar{\rho}_3^{(n-1)}$.
Of course, the boundaries will play some role too in this relation
\begin{multline} \label{eq:grossa}
  \bar{\rho}_3^{(n)} =
  \frac{1}{2^{n+2}-2} \left[ \Ki_L \left( \rho^{(n-1)}_{A,1,2} \right) + 
  \Ki_R \left( \rho^{(n-1)}_{2^{n+1}-1, 2^{n+1}, A'} \right) \right]
  \\ + 
  \left(1 - \frac{1}{2^{n+1} -1}\right) \cdot \Di \left(\bar{\rho}_3^{(n-1)} \right),
\end{multline}
(here $\Di$ is the average of $\Di_L$ and $\Di_R$). 
This equation shows the contributions of both bulk and edge terms; nevertheless,
when we approach the thermodynamical limit $n \to \infty$, the contribution
of the first two terms vanishes in every norm, because any density matrix
has trace norm bounded by one and CPT maps are contractive.

This means that the extensive influence of the boundary upon the lattice
grows \emph{slower} than the size of the system, a physical sounding and known property.
To quantitatively describe such behavior,
we  compute the (total) energy associated with the block of the first $2^\tau-1$ spins near a boundary,
say the left one. In our notation this corresponds to 
\begin{equation} \label{eq:estens}
 E^{(n)}_{1 \ldots 2^{\tau}-1} =
 \trace\left[ H_3 \cdot \sum_{j = 1}^{2^\tau-1} \rho^{(n)}_{j,j+1,j+2} \right].
\end{equation}
Exploiting the usual formalism of level-growing CPT maps, we can
rewrite the sum in~\eqref{eq:estens} as
\begin{equation}
 \sum_{j = 1}^{2^\tau-1} \rho^{(n)}_{j,j+1,j+2}
 = \sum_{p = 0}^{\tau - 1} 
 \, 2^p \cdot \Di^p \circ \Ki_L \circ \Bi_L^{\tau - p -1} \left( \rho^{(n-\tau)}_{A,1,2}\right).
\end{equation}
Now, we can successfully approach the thermodynamical limit while keeping $\tau$ fixed.
Recalling that $\rho^{f}_{A,1,2}$ is the fixed point of $\Bi_L$, we obtain
\begin{equation}
 E^{\tm}_{1 \ldots 2^{\tau}-1} = \trace \left[ H_3 \cdot \sum_{p = 0}^{\tau - 1} 
 \, 2^p \; \Di^p \circ \Ki_L \left( \rho^{f}_{A,1,2}\right) \right].
\end{equation}
As expected, the result diverges for $\tau \to \infty$ since the
series is made of terms growing in trace norm.
To explicitly estimate how this quantity \emph{deviates} from
its corresponding value in the bulk as $\tau$ grows we evaluate the following quantity  
\begin{multline}
 \Delta E^{\tm}_{1 \ldots 2^{\tau}-1} =  - (2^{\tau}- 1) \;\trace \left[ H_3 \cdot \rho^f_3 \right]
 + \\ +
 \trace \left[ H_3 \cdot \sum_{p = 0}^{\tau - 1} 
 \, 2^p \; \Di^p \circ \Ki_L \left( \rho^{f}_{A,1,2}\right) \right]
 = \\ =
 \trace \left[ H_3 \cdot \sum_{p = 0}^{\tau - 1} 
 \, 2^p \; \Di^p \left\{ \Ki_L \left( \rho^{f}_{A,1,2} \right) - \rho^f_3 \right\} \right].
\end{multline}
Notice that in this case the map $\Di$ applies to a traceless
operator, therefore if we decompose the argument in a basis of generalized eigenvectors
for $\Di$, it must have null component over the unique state of eigenvalue one.
As a result the boundary contribution to the ground state energy has the form
\begin{equation} \label{eq:diverg}
 \Delta E^{\tm}_{1 \ldots \infty} = \sum_{\kappa_D \neq 1} \; \sum_{p=0}^{\infty}
 ( 2 \kappa_{\Di} )^{p} \cdot g_{\kappa_{\Di}}(p),
\end{equation}
where $g(\cdot)$ is a polynomial in its main argument.
Looking at Eq.  \eqref{eq:diverg}, we notice that the inner sum diverges for any
eigenvalue $\kappa_{\Di}$ of $\Di$ greater or equal to $1/2$, unless the $g$ are identically zero
for such values of $\kappa_{\Di}$. 
In general this will happen
when $\rho^f_3 - \Ki_L(\rho^f_{A,1,2})$ has null component over any
generalized eigenspace whose $\kappa_{\Di} \geq 1/2$.

Interestingly enough the capability of such deviation to diverge is another
manifestation that the MERA states of Fig.~\ref{fig:chibera} are critical: indeed only 
the integral of a power-law function can diverge, while for an exponential decaying correlation function
the integral of the corresponding action is always finite.

\section{Conclusions} 
\label{sec:con}

In this paper we exploited the properties of MERA to describe boundary critical phenomena. We considered the 
case of a one-dimensional critical system with a boundary.  To this end we modified the local structure of the MERA 
at the boundary to account for more flexibility in its description.  

Besides showing how to compute local observables,  we achieved two main results. 
First of all we showed, as predicted by boundary conformal field theory, that  the critical 
exponents associated to the decay of the one-point function (as a function of the distance from the  boundary) is always 
half of the one of the bulk two-point correlation function corresponding to the same scaling operator.  
Secondly we compute the boundary corrections to the ground state energy and determined its scaling behavior. 
As in the bulk case, also in the presence of the boundary, most of the critical properties are determined solely by the eigenvalues of the 
MERA transfer matrix. 

A remarkable feature of treating boundary critical phenomena with MERAs is that it is enough to consider uniform tensor 
network. This is the result of the scale invariance of the underlying tensor network which holds also in the presence of 
boundaries.  In addition to the practical advantage in the numerical simulations, this observation  further clarifies the 
properties of the MERA. 
It is worth noticing that such property (that is at the basis of the effectiveness of a bounded MERA) is physically equivalent to the 
fact that, in boundary critical phenomena, the operator content of the bulk is not influenced by the boundary \cite{bcs}, suggesting 
that maybe the connection between MERA and general renormalization group theory is deeper than what nowadays understood.

One dimensional systems display the peculiar feature that the boundary can be critical only when also the bulk is. 
This is not the case in higher dimensions \cite{bcs}, where we can have a critical boundary in a gapped system, resulting in a 
richer scenario for the boundary-bulk phase diagram.  
This richness will reflect in the possibility to have different compositions of tensor structure. 
In this paper we considered a matrix product state (at the border) connected to a MERA. 
It is easy to imagine that, to describe critical surfaces in a non-critical bulk, different compositions of tensor 
networks are required.

During the writing of this work a paper by G. Evenbly {\em et al.}
appeared on the archive discussing boundary critical phenomena using MERA, see Ref.~\cite{BOU}.

 We acknowledge fruitful discussions with M. Campostrini, S. Montangero and  M. Rizzi,  and financial support from  
 FIRB-RBID08B3FM  and the National Research Foundation and Ministry of Education 
Singapore.


\begin{thebibliography}{99}
\bibitem{footbcco}
	Notice that allowing the boundary elements to change along the renormalization flow corresponds to introduce a 
	boundary condition changing operator \cite{cardy-05} in the language of CFT.
\bibitem{MOTA}
	The three-side observable is  the simplest configuration we can consider since the three-site causal cone is the
	smallest stable for a MERA-like bulk~\cite{mera}.
\end{thebibliography}
\end{document}